\documentclass[aip,apl,reprint,showpacs,superscriptaddress]{revtex4-1}
\usepackage[T1]{fontenc}
\usepackage{times}
\usepackage{graphicx,color}
\usepackage{amsfonts,amsmath,amssymb,amsbsy}
\usepackage[colorlinks=true, linkcolor=blue, citecolor=blue, urlcolor=blue]{hyperref}
\def\emph#1{\textcolor{blue}{#1}}

\begin{document}

\title{{Dynamics of a magnetic skyrmionium driven by spin waves}}

\author{Sai Li}
\thanks{These authors contributed equally to this work.}
\affiliation{Fert Beijing Institute, BDBC, and School of Electronic and Information Engineering, Beihang University, Beijing 100191, China}
\affiliation{School of Science and Engineering, The Chinese University of Hong Kong, Shenzhen 518172, China}
\affiliation{Shenyuan Honors College, Beihang University, Beijing 100191, China}

\author{Jing Xia}
\thanks{These authors contributed equally to this work.}
\affiliation{School of Science and Engineering, The Chinese University of Hong Kong, Shenzhen 518172, China}

\author{Xichao Zhang}
\affiliation{School of Science and Engineering, The Chinese University of Hong Kong, Shenzhen 518172, China}

\author{Motohiko Ezawa}
\affiliation{Department of Applied Physics, The University of Tokyo, Hongo 7-3-1, Tokyo 113-8656, Japan}

\author{Wang Kang}
\affiliation{Fert Beijing Institute, BDBC, and School of Electronic and Information Engineering, Beihang University, Beijing 100191, China}

\author{Xiaoxi Liu}
\affiliation{Department of Electrical and Computer Engineering, Shinshu University, 4-17-1 Wakasato, Nagano 380-8553, Japan}

\author{\\ Yan Zhou}
\email[Electronic mail:]{zhouyan@cuhk.edu.cn}
\affiliation{School of Science and Engineering, The Chinese University of Hong Kong, Shenzhen 518172, China}

\author{Weisheng Zhao}
\email[Electronic mail:]{weisheng.zhao@buaa.edu.cn}
\affiliation{Fert Beijing Institute, BDBC, and School of Electronic and Information Engineering, Beihang University, Beijing 100191, China}

\begin{abstract}
The magnetic skyrmionium is a skyrmion-like structure but carries a zero net skyrmion number, which can be used as a building block for non-volatile information processing devices. Here, we study the dynamics of a magnetic skyrmionium driven by propagating spin waves. It is found that the skyrmionium can be effectively driven into motion by spin waves showing tiny skyrmion Hall effect, of which the mobility is much better than that of the skyrmion at the same condition. We also show that the skyrmionium mobility depends on the nanotrack width and damping coefficient, and can be controlled by an external out-of-plane magnetic field. Besides, we demonstrate the skyrmionium motion driven by spin waves is inertial. Our results indicate that the skyrmionium is a promising building block for building spin-wave spintronic devices.
\end{abstract}

\date{23 March 2018}
\keywords{skyrmion, skyrmionium, spin wave, spintronics, micromagnetics}
\pacs{75.60.Ch, 75.70.Kw, 75.78.-n, 12.39.Dc}

\maketitle


Recently, there has been considerable interest in studying the topological objects in magnetic materials such as skyrmions and skyrmioniums~\cite{Wiesendanger_Review2016,Wanjun_PHYSREP2017}, because they can be used to carry information in a non-volatile manner~\cite{Iwasaki_NCOMMS2013,Sampaio_NNANO2013} and thus are promising for a wide range of applications~\cite{Kang_PIEEE2016,Fert_NATREVMAT2017,Bhatti_MAT2017}. The presence of skyrmions in chiral magnets was theoretically predicted~\cite{Roszler_NATURE2006} and has been experimentally verified in magnetic materials with the Dzyaloshinskii-Moriya interaction (DMI)~\cite{Muhlbauer_SCIENCE2009,Yu_NATURE2010}. It has also been reported that skyrmions can be used for building several types of information processing devices including racetrack-type memories~\cite{Sampaio_NNANO2013,Iwasaki_NCOMMS2013,Xichao_SREP2015A,Kang_IEEEEDL2016,Chen_APL2017} and logic gates~\cite{Xichao_SREP2015B}. Recent studies further show the potential of using skyrmions in reservoir computing systems~\cite{Prychynenko_PRAPPL2018} and bio-inspired spintronic applications~\cite{Yangqi_IOP2017,Lisai_IOP2017}.

On the other hand, the magnetic skyrmionium is also a magnetic spin texture stabilized by the DMI~\cite{Bogdanov_JMMM1999,Finazzi_PRL2013,LiuY_PRB2015,LiuQF_AIP2015,Beg_SREP2015,Xichao_PRB2016C,Mulkers_PRB2016,Komineas_PRB2015A,Komineas_PRB2015B,Zheng_PRL2017,Fujita_PRB2017A,Fujita_PRB2017B,Shilei_NANOLETT2018}. It can be topologically viewed as a combination of two skyrmions carrying opposite skyrmion numbers~\cite{Xichao_PRB2016C,Zheng_PRL2017}. Namely, the skyrmionium is a skyrmion-like structure but has a zero net skyrmion number.
It is worth mentioning that the skyrmionium is also referred to as the $3\pi$-skyrmion~\cite{Rohart_PRB2013} and target skyrmion~\cite{Leonov_EPJ2014,Beg_SREP2015,Beg_PRB2017,Carey_APL2016,Pepper_JAP2018}.
Skyrmioniums can also be used as building blocks for information processing devices~\cite{Xichao_PRB2016C}, while their dynamics and response to external stimuli are different from those of skyrmions. For example, the dynamics of a skyrmionium driven by spin transfer torques has been studied~\cite{Xichao_PRB2016C,Komineas_PRB2015B}, which shows the skyrmionium has no skyrmion Hall effect~\cite{Zang_PRL2011,Wanjun_NPHYS2017,Litzius_NPHYS2017}. Recently, experimental and theoretical studies suggested it is possible to create skyrmioniums by using ultra-fast laser pulses~\cite{Finazzi_PRL2013,Fujita_PRB2017A,Fujita_PRB2017B}. Besides, the skyrmionium dynamics driven by applied magnetic fields has also been theoretically studied~\cite{Komineas_PRB2015A} and experimentally observed~\cite{Zheng_PRL2017,Shilei_NANOLETT2018}. However, the dynamics of skyrmioniums induced by spin waves still remain elusive, although it has been examined for skyrmions~\cite{Walliser_PRB2000,Kong_PRL2013,Lin_PRL2014,Iwasaki_PRB2014,Xichao_NANOTECH2015,Xichao_NJP2017}.

In this paper, we theoretically and numerically study the dynamics of a skyrmionium driven by spin waves in a magnetic thin film with interface-induced DMI, which is in stark contrast to that of the skyrmion. We also investigate the motion of a skyrmionium in a narrow nanotrack driven by spin waves propagating along the nanotrack. We find that the skyrmionium can be effectively driven into motion by spin waves and reach higher speed in compared with the skyrmion. The skyrmionium mobility driven by spin waves can also be controlled by an external magnetic field and is subject to other parameters.

%

\begin{figure}[t]
\centerline{\includegraphics[width=0.48\textwidth]{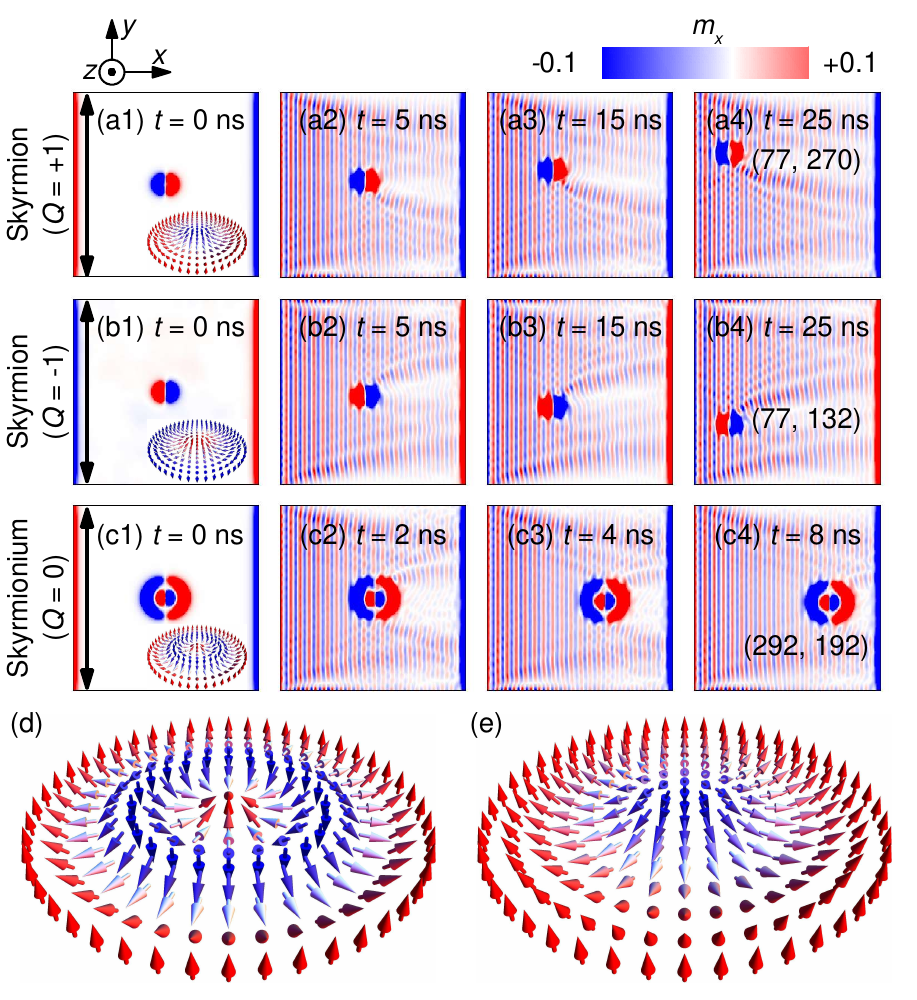}}
\caption{%
Snapshots of skyrmions and skyrmionium driven by spin waves in thin films at selected times before reaching the edge.
(a) A skyrmion with $Q=+1$.
(b) A skyrmion with $Q=-1$, i.e., an antiskyrmion.
(c) A skyrmionium with $Q=0$.
(d) Schematic of a skyrmionium.
(e) Schematic of a skyrmion.
The insets in (a1), (b1) and (c1) show the spin textures of the skyrmion, antiskyrmion, and skyrmionium.
The coordinates in (a4), (b4) and (c4) indicate the locations of the skyrmion, antiskyrmion, and skyrmionium.
For all cases, an oscillating magnetic field is applied at the left film edge ($x<15$ nm), exciting spin waves propagating toward the $+x$ direction.
The color scale represents the in-plane spin component $m_{x}$. Here $\alpha=0.02$.
}
\label{FIG1}
\end{figure}

We perform the simulation using the Object Oriented MicroMagnetic Framework (OOMMF)~\cite{OOMMF}, which solves the spin dynamics based on the Landau-Lifshitz-Gilbert equation~\cite{Gilbert_1955}. The detailed methods are given in Ref.~\onlinecite{Xichao_PRB2016C}.
The default magnetic material parameters are adopted from Refs.~\onlinecite{Xichao_PRB2016C,Xichao_NJP2017}: Heisenberg exchange constant $A=15$ pJ m$^{-1}$, DMI constant $D=3.5$ mJ m$^{-2}$, perpendicular magnetic anisotropy (PMA) constant $K=0.8$ MJ m$^{-3}$, and saturation magnetization $M_{\text{S}}=580$ kA m$^{-1}$. The mesh size is $2$ nm $\times$ $2$ nm $\times$ $1$ nm.
In experiments, these material parameters correspond to the ultrathin ferromagnet/heavy metal heterostructure system, where the DMI is induced at the interface.
The skyrmion number is defined as
$Q=-(1/4\pi) \int\boldsymbol{m}\cdot(\partial_{x}\boldsymbol{m}\times\partial_{y}\boldsymbol{m})dxdy$
in this paper, where $\boldsymbol{m}$ being the reduced magnetization and the minus sign ensures that the skyrmion with a spin-down core (pointing along the $-z$ direction) has a positive skyrmion number (cf. Fig.~\ref{FIG1}).

First, we put a skyrmion with $Q=+1$, a skyrmion with $Q=-1$, and a skyrmionium with $Q=0$ at the center of a magnetic film ($400$ nm $\times$ $400$ nm $\times$ $1$ nm) with the Gilbert damping coefficient $\alpha=0.02$, respectively, which are relaxed to metastable states. In this work, we refer to the skyrmion with $Q=-1$ as an antiskyrmion for simplicity.
Note that the out-of-plane spin structure of the antiskyrmion here differs from that of the skyrmion, while an antiskyrmion can also have different in-plane spin structure compared to the skyrmion~\cite{Xichao_SREP2015B,Koshibae_NCOMMS2016}.
We then apply an oscillating magnetic field $\boldsymbol{H}=H_\text{a}\sin⁡(2\pi ft)\hat{y}$ at the left edge of the film within a narrow area of width $15$ nm, where the field amplitude $H_\text{a} = 1000$ mT and frequency $f = 200$ GHz.
Smaller amplitude and frequency can also be used, but the driving force decreases with decreasing amplitude/frequency~\cite{Xichao_NANOTECH2015,Xichao_NJP2017}.
The oscillating magnetic field produces a flow of spin waves propagating toward the $+x$ direction with the wave vector $\boldsymbol{q}=q\hat{x}$, which diminishes progressively in the $x$ dimension because of the damping effect.

Figure~\ref{FIG1} shows top-view snapshots of skyrmion, antiskyrmion, and skyrmionium driven by spin waves at selected times. The interaction between the skyrmion (antiskyrmion) and spin waves is significantly different from that between the skyrmionium and spin waves, which results in the motion of the skyrmion and skyrmionium in opposite directions.
As shown in Fig.~\ref{FIG1}(a), the incident spin waves are obviously scattered by the skyrmion, leading to a backward motion of the skyrmion with a skew angle $\phi_{+}$. For the antiskyrmion in Fig.~\ref{FIG1}(b), the spin-wave-antiskyrmion scattering leads to the backward motion of the antiskyrmion with a skew angle $\phi_{-}$, which has the same magnitude but opposite sign to $\phi_{+}$. The scattering processes for the skyrmion and antiskyrmion are symmetric with respect to the horizontal axis through the skyrmion center. The skyrmion and antiskyrmion basically move toward the spin-wave-current source and can reach the left film edge, which agree well with the results in Refs.~\onlinecite{Schutte_PRB2014A,Schutte_PRB2014B,Iwasaki_PRB2014,Xichao_NJP2017}.

Interestingly, as the skyrmionium ($Q=0$) is composed of a skyrmion ($Q=+1$) and an antiskyrmion ($Q=-1$) (cf. Ref.~\onlinecite{Xichao_PRB2016C}), the scattering of spin waves by the skyrmionium can be regarded as a superposition of those of a skyrmion and an antiskyrmion, however, it does not result in the backward motion of the skyrmionium. In Fig.~\ref{FIG1}(c), it shows that the skyrmionium moves along the direction of the spin-wave current and can reach the right film edge, which behaves like a Newtonian particle driven by the moment transfer of the spin wave to it.

\begin{figure}[t]
\centerline{\includegraphics[width=0.48\textwidth]{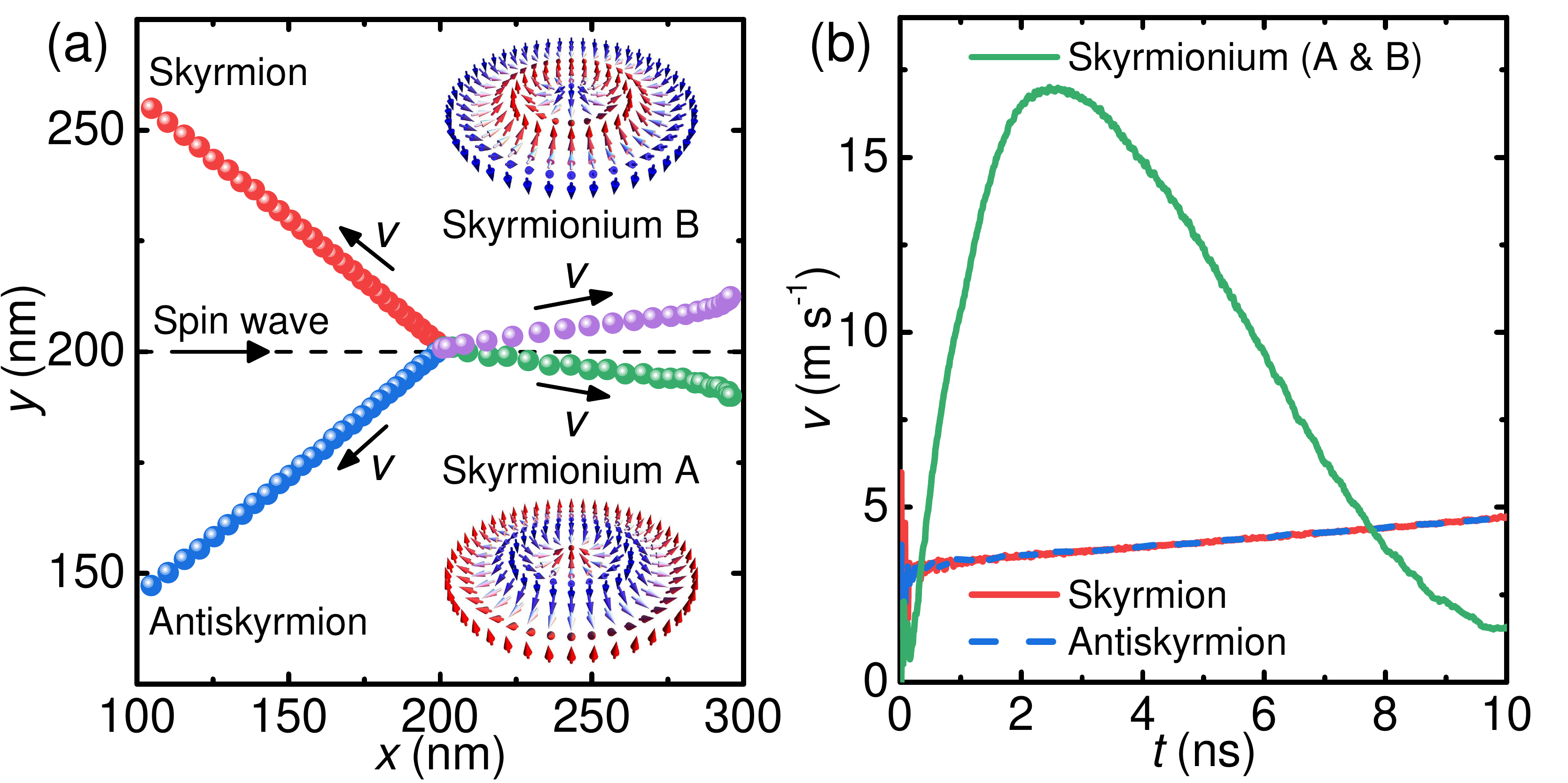}}
\caption{%
(a) The trajectories of skyrmions and skyrmionium driven by spin waves in thin films, corresponding to Fig.~\ref{FIG1}. The dot represents the center of the skyrmion or skyrmionium. Two skyrmionium structures are considered (cf. insets): the skyrmionium A is composed of the outer skyrmion and the inner antiskyrmion, while the skyrmionium B is composed of the outer antiskyrmion and the inner skyrmion.
(b) The velocities of skyrmions and skyrmioniums as functions of time.
}
\label{FIG2}
\end{figure}

\begin{figure}[t]
\centerline{\includegraphics[width=0.45\textwidth]{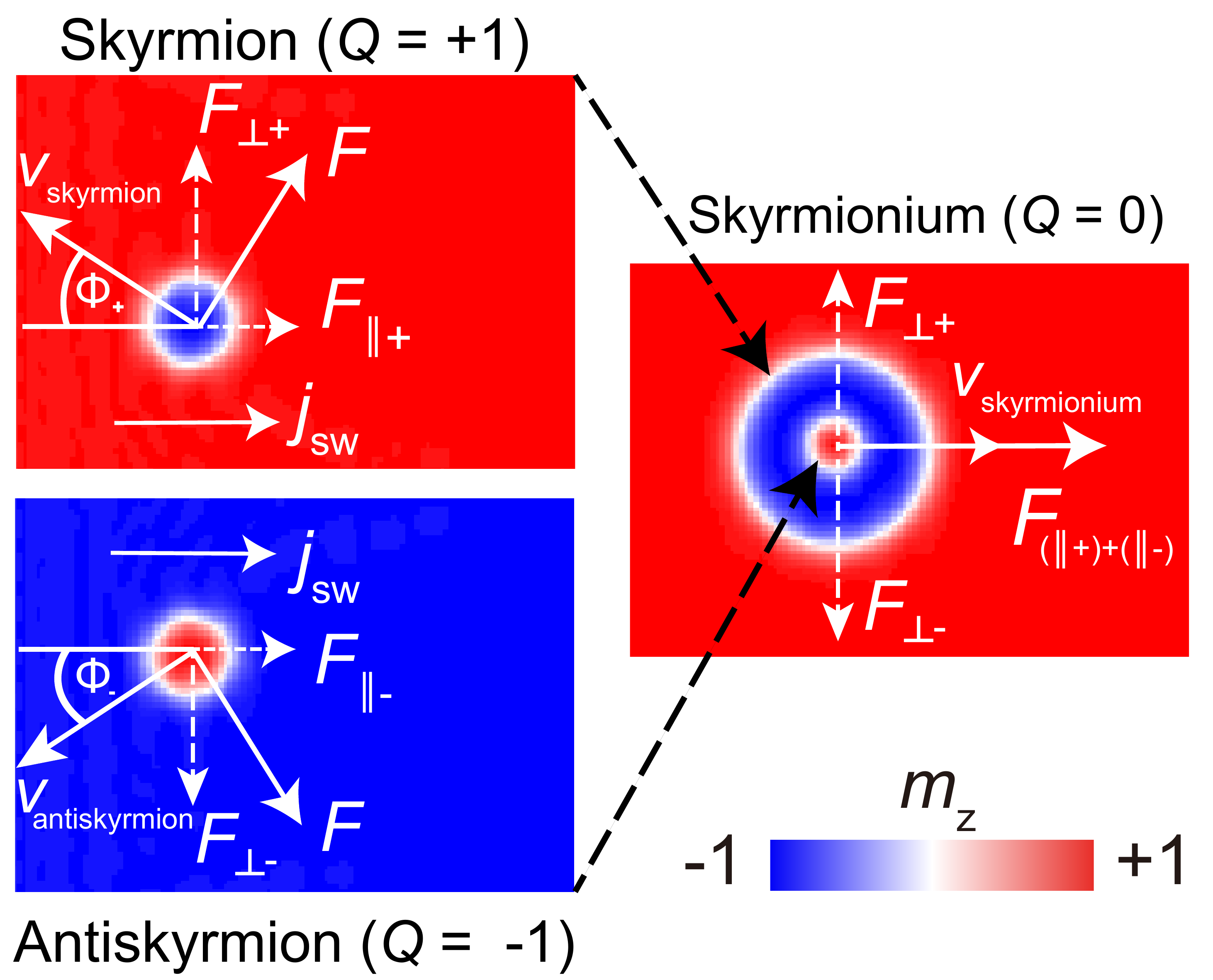}}
\caption{%
Illustration of the directions of the spin-wave currents, spin-wave driving forces (cf. Eq.~\ref{eq:force}), and the velocities for the skyrmion, antiskyrmion, and skyrmionium. ``+'' and ``-'' stand for the skyrmion and antiskyrmion, respectively.
}
\label{FIG3}
\end{figure}

The trajectories of the skyrmion, antiskyrmion, and skyrmionium driven by spin waves in the film are given in Fig.~\ref{FIG2}(a), where the time spans are corresponding to the Fig.~\ref{FIG1}. It can be seen that the skyrmion shows a negative longitudinal motion with a positive transverse motion, while the antiskyrmion shows both negative longitudinal and transverse motion.
For the skyrmionium, two possible structures are considered (cf. Fig.~\ref{FIG2}(a) insets): the one is composed of the outer skyrmion and the inner antiskyrmion, and the other one is composed of the outer antiskyrmion and the inner skyrmion.
The motion direction of the skyrmionium is almost along the positive longitudinal direction. Note that the skyrmionium has a very small transverse velocity, which we will discussed later. It is worth mentioning that once the skyrmion and skyrmionium reach the film edge, they will experience the edge repulsive force (cf. Ref.~\onlinecite{Xichao_NJP2017}). In this work, we only focus on the dynamics of the skyrmionium before reaching the sample edge.

The corresponding velocities of the skyrmion and skyrmionium are given in Fig.~\ref{FIG2}(b). The Hall angles of the skyrmion and antiskyrmion are the same, and their velocity magnitudes are identical. The skyrmion velocity also increases with time, as the driving force provided by spin waves increases with decreasing distance between the skyrmion and the spin wave source.
On the other hand, the skyrmionium velocity rapidly increases once the spin wave reaches it. It moves much faster than the skyrmion. The skyrmionium velocity reaches $16.4$ m s$^{-1}$ at $t=2$ ns, while the skyrmion velocity equals only $3.6$ m s$^{-1}$ at $t=2$ ns.
However, the skyrmionium decelerates after $t=2.5$ ns because it is close to the right edge of the film.
Note that the spin wave driving force also decreases with increasing distance between the skyrmionium and the spin wave source. Nevertheless, for the time span of $t<2$ ns, the motion of skyrmionium can be seen as being solely driven by spin waves.


We proceed to analyze the skyrmionium dynamics using the Thiele approach (cf. Refs.~\onlinecite{Thiele_PRL1973,Schutte_PRB2014A,Schutte_PRB2014B,Xichao_NANOTECH2015,Xichao_NJP2017}) assuming the skyrmionium is a rigid object (i.e., the distortion is negligible) composed of a skyrmion and an antiskyrmion with identical size, and explain why the skyrmionium is almost moving along the direction of the spin-wave current.
The skyrmionium position $\boldsymbol{R}$ can be described within the Thiele equation given as
\begin{equation}
\boldsymbol{G}\times\boldsymbol{\dot{R}}+\alpha\mathcal{D}\cdot\boldsymbol{\dot{R}}=\boldsymbol{F}(\boldsymbol{R}),
\label{eq:Thiele}
\end{equation}
where we take partial derivative of $\boldsymbol{R}$ with respect to time $t$, obtaining $\boldsymbol{\dot{R}}$ as the velocity of skyrmionium $\boldsymbol{v}$, $\alpha$ is the damping constant, $\mathcal{D}$ is the dissipative force tensor, and $\boldsymbol{G}=G\hat{z}$ is the gyromagnetic coupling vector, which can be expressed as
\begin{equation}
\boldsymbol{G}=(0,0,4\pi Q),
\label{eq:G}
\end{equation}
where $Q$ is the skyrmion number. For the skyrmionium with $Q=0$, $\boldsymbol{G}=\boldsymbol{0}$.
When the skyrmionium is far away from film edges, it only experiences the driving force provided by spin waves.
The total force acting on the skyrmionium is $\boldsymbol{F}(\boldsymbol{R})=\boldsymbol{F}_{\text{sw}}(\boldsymbol{R})$, resulting from the momentum transfer between the spin-wave current and the skyrmionium, which reads~\cite{Xichao_NJP2017}
\begin{equation}
\boldsymbol{F}_{\text{sw}}(\boldsymbol{R})=je^{\frac {\boldsymbol{r}\cdot\hat{q}}{\boldsymbol{L}_\text{sw}}}q(\sigma_{\parallel}\hat{q}+\sigma_{\perp}(\hat{z}\times\hat{q})),
\label{eq:force}
\end{equation}
where $j>0$ is the spin-wave current density, and from wave vector $\boldsymbol{q}=q\hat{x}$, we can get $\hat{q}=\hat{x}$. The exponential term is the approximation form, indicating that the decay of the spin-wave current on the scale relating to the Gilbert damping, $\frac{1}{\boldsymbol{L}_\text{sw}}\approx\alpha\sqrt{\frac{m}{2\hbar}2\pi f}$, where $m$, $f$ are the spin wave mass and frequency.
It can be seen from Eq.~\ref{eq:force} that the spin wave force acting on the skyrmionium can be divided into two parts, namely, one is the force $\boldsymbol{F}_{\parallel}$ along the spin wave vector $\hat{x}$ determined by $\sigma_{\parallel}$ and the other one is $\boldsymbol{F}_{\perp}$ perpendicular to the wave vector $\hat{x}$ governed by $\sigma_{\perp}$. The longitudinal and transverse cross sections are expressed as~\cite{Schutte_PRB2014A,Schutte_PRB2014B}
\begin{equation}
\dbinom{\sigma_{\parallel}(\epsilon)}{\sigma_{\perp}(\epsilon)}=\int_{0}^{2\pi}d\chi\dbinom{1-cos\chi}{-sin\chi}\frac{d\sigma(\epsilon)}{d\chi},
\label{eq:sigma}
\end{equation}
where $\chi$ is the skew scattering angle of the spin wave, $d\sigma(\epsilon)/d\chi$ is the differential cross section obtained by $\left|f(\chi)\right|^{2}$. Thus, we find that $\sigma_{\parallel}$ and $\sigma_{\perp}$ are dependent on the skew scattering angle $\chi$.
With this analysis, for the skyrmion and antiskyrmion, the longitudinal cross sections $\sigma_{\parallel}(\phi_{+})=\sigma_{\parallel}(\phi_{-})$ and the transverse cross sections $\sigma_{\perp}(\phi_{+})=-\sigma_{\perp}(\phi_{-})$, resulting from $\phi_{+}=-\phi_{-}$.
The transverse and longitudinal forces for the skyrmion and antiskyrmion are illustrated in Fig.~\ref{FIG3}, which are consistent with the numerical results in Fig.~\ref{FIG2}.

As the skyrmionium is composed of a skyrmion and an antiskyrmion, the total effective force (cf. Fig.~\ref{FIG3}) acting on the skyrmionium includes the joint forces from the skyrmion and antiskyrmion, where the transverse force of skyrmionium is equal to zero under the assumption that the skyrmion size and antiskyrmion size are identical, and the longitudinal force is the sum of $\boldsymbol{F}_{\parallel+}$ and $\boldsymbol{F}_{\parallel-}$. Since the skyrmionium is only driven by the force $\boldsymbol{F}_{\parallel}$ along the $\hat{x}$ direction, we can obtain 
\begin{equation}
v_{x}\cdot\alpha\mathcal{D}=\boldsymbol{F}_{\text{sw}}(\boldsymbol{R})=je^{\frac {\boldsymbol{r}\cdot\hat{q}}{\boldsymbol{L}_\text{sw}}}\left|q\right|^{2}\sigma_{\parallel}\cdot\hat{x},
\label{eq:skyrmionium}
\end{equation}
which gives a good interpretation of the simulation results, namely, in the beginning of the simulation the longitudinal spin-wave driving force along the $\hat{x}$ direction is very strong so that the skyrmionium moves forward with a $\hat{x}$-direction velocity.
The longitudinal component of the velocity is attributed to the longitudinal scattering section of the skyrmionium, simplified as
\begin{equation}
v_{x}=\frac{j}{\alpha\mathcal{D}}e^{\frac {\boldsymbol{r}\cdot\hat{q}}{\boldsymbol{L}_\text{mag}}}\left|q\right|^{2}\sigma_{\parallel}\cdot\hat{x}.
\label{eq:vx}
\end{equation}
It should be noted that the skyrmionium is actually composed of a skyrmion and an antiskyrmion with different sizes, which means the transverse forces acting on the skyrmionium do not offset each other and will result in a small transverse motion of the skyrmionium. The transverse motion direction depends on the internal structure of the skyrmionium (cf. Fig.~\ref{FIG2}). The dynamics of the skyrmion in the internal structure of the skyrmionium can also be independently described by introducing an additional force term to Eq.~\ref{eq:Thiele}, which is similar to the spin-wave-driven motion of skyrmion along the edge, i.e., experiencing the edge repulsive force (cf. Ref.~\onlinecite{Xichao_NJP2017}). Besides, it is worth mentioning that the skyrmionium driven by spin transfer torques shows no skyrmion Hall effect (cf. Ref.~\onlinecite{Xichao_PRB2016C}).

\begin{figure}[t]
\centerline{\includegraphics[width=0.50\textwidth]{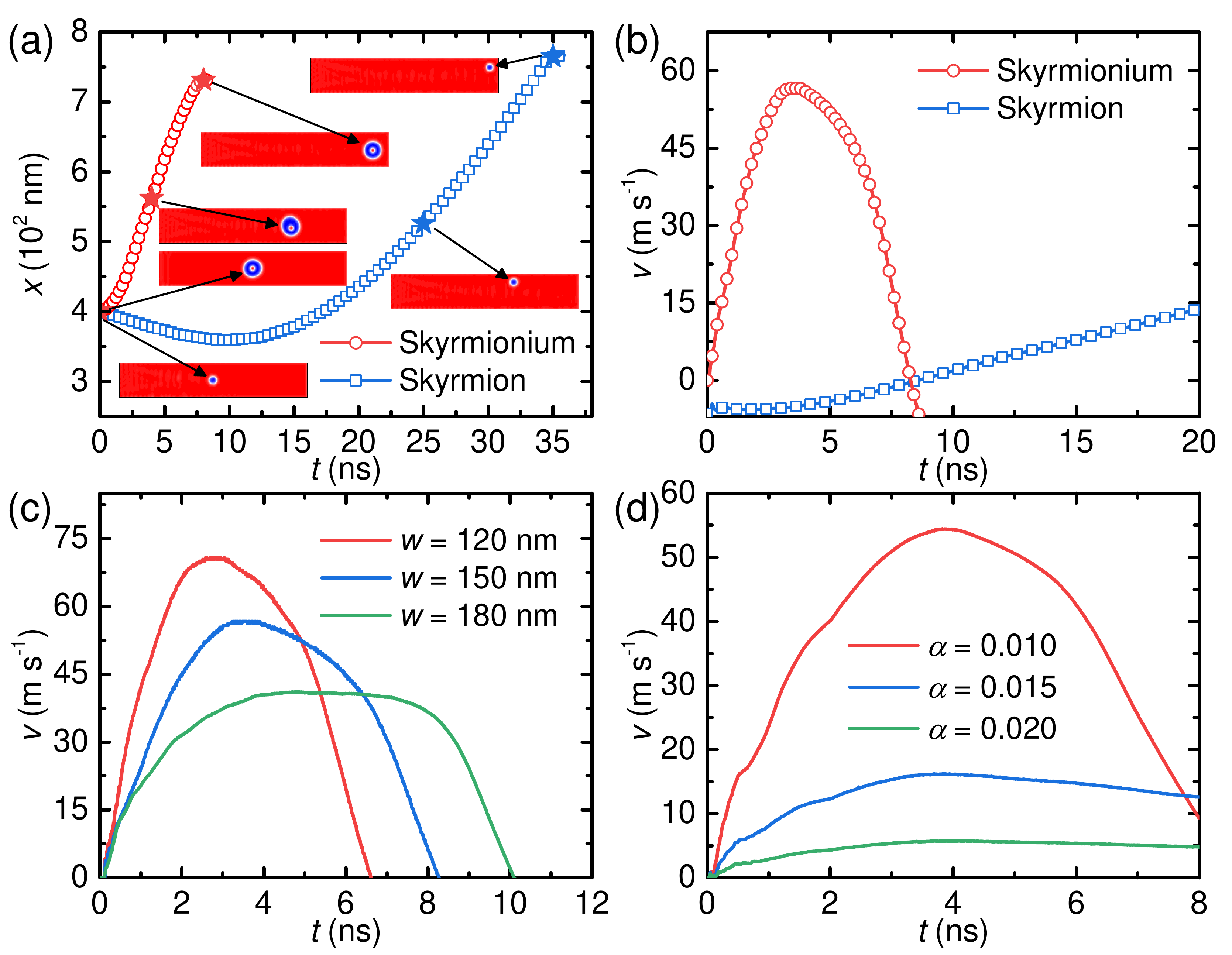}}
\caption{%
(a) The longitudinal coordinates ($x$) versus time for the skyrmionium and skyrmion driven by spin waves in a nanotrack ($800$ nm $\times$ $150$ nm $\times$ $1$ nm). The insets show top-view snapshots at selected times. Here $\alpha=0.01$.
(b) The velocities versus time corresponding to (a).
(c) The velocity versus time of the skyrmionium for different nanotrack widths. Here $\alpha=0.01$.
(d) The velocity versus time of the skyrmionium for different damping coefficients. Here $w=150$ nm.
}
\label{FIG4}
\end{figure}

\begin{figure}[t]
\centerline{\includegraphics[width=0.50\textwidth]{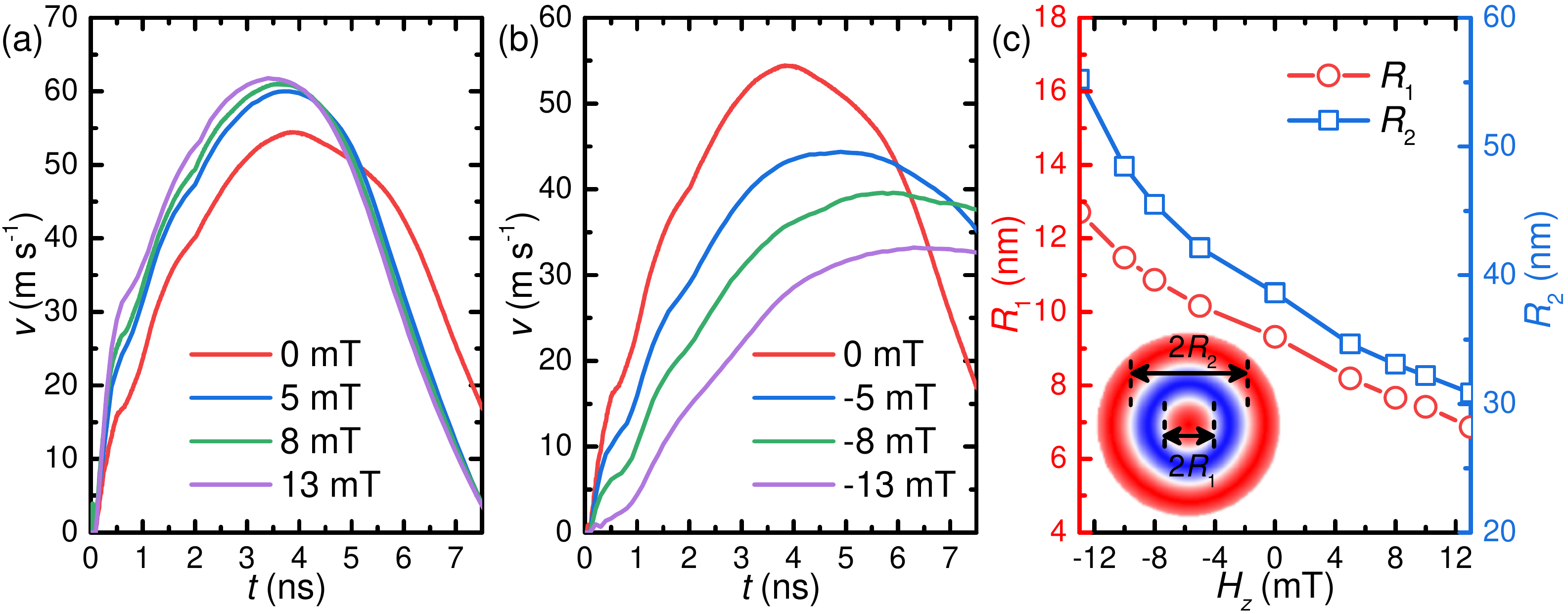}}
\caption{%
(a) The velocity versus time of the skyrmionium driven by spin waves in a nanotrack at $H_{z} > 0$.
(b) The velocity versus time of the skyrmionium driven by spin waves in a nanotrack at $H_{z} < 0$.
(c) The inner radius ($R_1$) and outer radius ($R_2$) of the skyrmionium as functions of $H_z$.
The inset illustrates the definitions of $R_1$ and $R_2$.
Here, the nanotrack size is $800$ nm $\times$ $150$ nm $\times$ $1$ nm and $\alpha=0.01$.
}
\label{FIG5}
\end{figure}


Because the skyrmionium driven by spin waves only has tiny skyrmion Hall effect, it can be used as a reliable information carrier in magnetic devices. Here, for the purpose of developing applications, we further numerically investigate the motion of a skyrmionium in the nanotrack driven by spin waves.

As shown in Fig.~\ref{FIG4}, we simulate the spin-wave-driven motion of a skyrmionium and a skyrmion in a nanotrack with length $l=800$ nm and width $w=150$ nm, respectively, where the spin wave source is applied at the left end of the nanotrack with parameters used above.
Figure~\ref{FIG4}(a) shows the longitudinal coordinate $x$ as a function of time for the skyrmionium and skyrmion, where the insets show the top-views of the nanotrack at selected times.
It can be seen that the skyrmionium driven by spin waves moves along the nanotrack toward the $+x$ direction with a slight distortion, which reaches the nanotrack end at $t=8$ ns. The skyrmionium accelerates and then decelerates, and can reach a maximum velocity of $57$ m s$^{-1}$ [cf. Fig.~\ref{FIG4}(b)].
In contrast, the skyrmion first moves toward the spin wave source (i.e., the $-x$ direction) and approaches the upper edge, then it moves along the upper edge toward the $+x$ direction. It moves to the nanotrack end at $t=35$ ns. The skyrmion velocity increases with time and only reaches a maximum velocity of $20$ m s$^{-1}$ [cf. Fig.~\ref{FIG4}(b)]. Due to the presence of the backward motion before reaching the upper edge, the skyrmion trajectory is subject to the nanotrack width, whereas there is no such a problem for the skyrmionium.

In order to further analyze the dependence of the skyrmionium dynamics driven by spin waves on different parameters, we simulate its motion in nanotracks with various $w$ and $\alpha$.
As shown in Fig.~\ref{FIG4}(c), the maximum velocity of the skyrmionium reduces with increasing nanotrack width, which means narrow nanotracks can be used for building high-speed skyrmionium shift devices. However, the skyrmionium will be distorted if the nanotrack width is smaller than the skyrmionium size. Hence, a nanotrack with a width slightly larger than the relaxed skyrmionium size can be used for piratical applications.
Figure~\ref{FIG4}(d) shows the skyrmionium velocity for different damping coefficients. It shows the skyrmionium mobility significantly decreases with increasing damping coefficient, which suggests the high-speed skyrmionium motion driven by spin waves can only be realized in low-damping materials.

Because the skyrmionium size can be controlled by an external magnetic field (cf. Ref.~\onlinecite{Xichao_PRB2016C}), we also apply an out-of-plane magnetic field to the nanotrack and observe its effect on the skyrmionium motion driven by spin waves.
It shows that the skyrmionium mobility increases with increasing $H_z$ [cf. Fig.~\ref{FIG5}(a)-(b)]. Namely, the maximum velocity of the skyrmionium increases with decreasing inner and outer radii of the skyrmionium, as the skyrmionium shrinks for $H_{z}>0$ mT and expands for $H_{z}<0$ mT [cf. Fig.~\ref{FIG5}(c)]. Note that the skyrmionium will be transformed to a skyrmion when ${H}_{z}<13$ mT and be distorted when $H_{z}>13$ mT.

\begin{figure}[t]
\centerline{\includegraphics[width=0.45\textwidth]{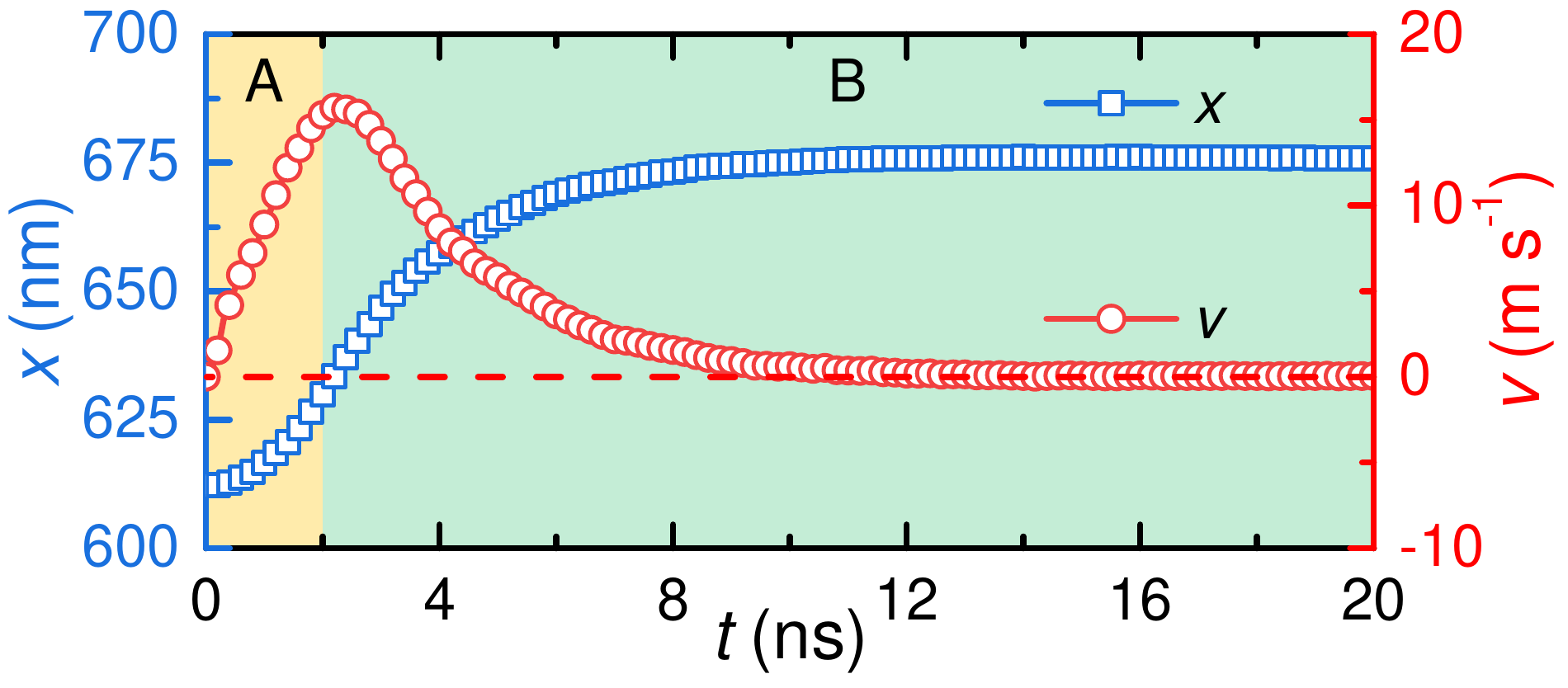}}
\caption{%
The longitudinal coordinates ($x$) and velocity as functions of time for the skyrmionium driven by spin waves in a nanotrack ($1200$ nm $\times$ $150$ nm $\times$ $1$ nm). The spin wave source is switched off at $t=2$ ns.
}
\label{FIG6}
\end{figure}

Last, as shown in Fig.~\ref{FIG6}, we examine the skyrmionium dynamics when the spin wave source is suddenly switched off. It can be seen that the skyrmionium is first driven into motion by propagating spin waves during the first $2$ ns [cf. scenario A in Fig.~\ref{FIG6}]. At $t=2$ ns, we switch off the spin wave source and the skyrmionium starts to decelerate when all spin waves are damped [cf. scenario B in Fig.~\ref{FIG6}]. Finally, the skyrmionium velocity reduced to zero and reaches a new location in the nanotrack. Therefore, it can be seen the skyrmionium motion driven by spin waves is inertial.
Namely, it can be regarded as a Newtonian quasi-particle.


In summary, we have studied the skyrmionium motion driven by propagating spin waves in both thin films and nanotracks. We find that the skyrmionium moves along the spin wave propagation direction with tiny skyrmion Hall effect, and can reach a much larger velocity than the skyrmion. In contrast, the skyrmion driven by spin waves will first show a backward motion toward the spin wave source. We also find the skyrmionium mobility depends on the nanotrack width and damping coefficient. Moreover, we demonstrate that the skyrmionium mobility can be tuned by apply an out-of-plane magnetic field. We also show the skyrmionium motion driven by spin waves is inertial. Our results may be useful for developing spintronic applications based on the skyrmionium driven by spin waves.


This work is supported by the National Natural Science Foundation of China (Grant Nos. 11574137, 61471015, 61501013 and 61571023), the Shenzhen Fundamental Research Fund (Grant Nos. JCYJ20160331164412545 and JCYJ20170410171958839), the President's Fund of CUHKSZ, the National Key Technology Program of China (Grant No. 2017ZX01032101), the International Mobility Projects (Grant Nos. B16001 and 2015DFE12880), the JSPS RONPAKU (Dissertation Ph.D.) Program, the Grants-in-Aid for Scientific Research from JSPS KAKENHI (Grant Nos. JP17K05490 and JP15H05854), and CREST, JST (Grant No. JPMJCR16F1).

 


\begin{thebibliography}{} 

\bibitem{Wiesendanger_Review2016} R. Wiesendanger, Nat. Rev. Mat. \textbf{1}, 16044 (2016).

\bibitem{Wanjun_PHYSREP2017} W. Jiang, G. Chen, K. Liu, J. Zang, S.~G. Velthuiste and A. Hoffmann, Phys. Rep. \textbf{704}, 1 (2017).

\bibitem{Iwasaki_NCOMMS2013} J. Iwasaki, M. Mochizuki and N. Nagaosa, Nat. Commun. \textbf{4}, 1463 (2013).

\bibitem{Sampaio_NNANO2013} J. Sampaio, V. Cros, S. Rohart, A. Thiaville and A. Fert, Nat. Nanotech. \textbf{8}, 839 (2013).

\bibitem{Fert_NATREVMAT2017} A. Fert, N. Reyren and V. Cros, Nat. Rev. Mat. \textbf{2}, 17031 (2017).

\bibitem{Bhatti_MAT2017} S. Bhatti, R. Sbiaa, A. Hirohata, H. Ohno, S. Fukami and S. N. Piramanayagam, Mater. Today. \textbf{20}, 530--548, (2017).

\bibitem{Kang_PIEEE2016} W. Kang, Y. Huang, X. Zhang, Y. Zhou and W. Zhao, Proceedings of the IEEE \textbf{104}, 2040--2061 (2016).

\bibitem{Roszler_NATURE2006} U.~K. R{\"o}{\ss}ler, A.~N. Bogdanov and C. Pfleiderer, Nature \textbf{442}, 797 (2006).

\bibitem{Muhlbauer_SCIENCE2009} S. M{\"u}hlbauer, B. Binz, F. Jonietz, C. Pfleiderer, A. Rosch, A. Neubauer, R. Georgii and P. B{\"o}ni, Science \textbf{323}, 915 (2009).

\bibitem{Yu_NATURE2010} X.~Z. Yu, Y. Onose, N. Kanazawa, J.~H. Park, J.~H. Han, Y. Matsui, N. Nagaosa and Y. Tokura, Nature \textbf{465}, 901 (2010).

\bibitem{Xichao_SREP2015A} X. Zhang, G.~P. Zhao, H. Fangohr, J.~P. Liu, W.~X. Xia, J. Xia and F.~J. Morvan, Sci. Rep. \textbf{5}, 7643 (2015).

\bibitem{Kang_IEEEEDL2016} W. Kang, C. Zheng, Y. Huang, X. Zhang, Y. Zhou, W. Lv and W. Zhao, IEEE Electron Device Lett. \textbf{37}, 924 (2016).

\bibitem{Chen_APL2017} X. Chen, W.Kang, D.Zhu, X. Zhang, N. Lei, Y. Zhang, Y. Zhou and W. Zhao, Appl. Phys. Lett. \textbf{111}, 202406 (2017).

\bibitem{Xichao_SREP2015B} X. Zhang, M. Ezawa and Y. Zhou, Sci. Rep. \textbf{5}, 9400 (2015).

\bibitem{Prychynenko_PRAPPL2018} D. Prychynenko, M. Sitte, K. Litzius, B. Kr{\"u}ger, G. Bourianoff, M. Kl{\"a}ui, J. Sinova and K. Everschor-Sitte, Phys. Rev. Appl. \textbf{9}, 014034 (2018).

\bibitem{Yangqi_IOP2017} Y. Huang, W. Kang, X. Zhang, Y. Zhou and W. Zhao, Nanotechnology \textbf{28}, 08LT02 (2017).

\bibitem{Lisai_IOP2017} S. Li, W. Kang, Y. Huang, X. Zhang, Y. Zhou and W. Zhao, Nanotechnology \textbf{28}, 31LT01 (2017).

\bibitem{Bogdanov_JMMM1999} A. Bogdanov and A. Hubert, J. Magn. Magn. Mater. \textbf{195}, 182 (1999).

\bibitem{Finazzi_PRL2013} M. Finazzi, M. Savoini, A.~R. Khorsand, A. Tsukamoto, A. Itoh, L. Du\`o, A. Kirilyuk, T. Rasing and M. Ezawa, Phys. Rev. Lett. \textbf{110}, 177205 (2013).

\bibitem{LiuY_PRB2015} Y. Liu, H. Du, M. Jia and A. Du, Phys. Rev. B \textbf{91}, 094425 (2015).

\bibitem{Beg_SREP2015} M. Beg, R. Carey, W. Wang, D. Cort{\'e}s-Ortu{\~n}o, M. Vousden, M.-A. Bisotti, M. Albert, D. Chernyshenko, O. Hovorka, R.~L. Stamps and H. Fangohr, Sci. Rep. \textbf{5}, 17137 (2015).

\bibitem{LiuQF_AIP2015} X. Liu, Q. Zhu, S. Zhang, Q. Liu and J. Wang, AIP Adv. \textbf{5}, 087137 (2015).

\bibitem{Xichao_PRB2016C} X. Zhang, J. Xia, Y. Zhou, D. Wang, X. Liu, W. Zhao and M. Ezawa, Phys. Rev. B \textbf{94}, 094420 (2016).

\bibitem{Mulkers_PRB2016} J. Mulkers, M.~V. Milo{\v{s}}evi{\'{c}} and B. Van~Waeyenberge, Phys. Rev. B \textbf{93}, 214405 (2016).

\bibitem{Komineas_PRB2015A} S. Komineas and N. Papanicolaou, Phys. Rev. B \textbf{92}, 064412 (2015).

\bibitem{Komineas_PRB2015B} S. Komineas and N. Papanicolaou, Phys. Rev. B \textbf{92}, 174405 (2015).

\bibitem{Zheng_PRL2017} F. Zheng, H. Li, S. Wang, D. Song, C. Jin, W. Wei, A. Kov\'acs, J. Zang, M. Tian, Y. Zhang, H. Du and R.~E. Dunin-Borkowski, Phys. Rev. Lett. \textbf{119}, 197205 (2017).

\bibitem{Fujita_PRB2017A} H. Fujita and M. Sato, Phys. Rev. B \textbf{95}, 054421 (2017).

\bibitem{Fujita_PRB2017B} H. Fujita and M. Sato, Phys. Rev. B \textbf{96}, 060407 (2017).

\bibitem{Shilei_NANOLETT2018} S. Zhang, F. Kronast, G. Laanvan~der and T. Hesjedal, Nano Lett. \textbf{18}, 1057--1063 (2018). 

\bibitem{Rohart_PRB2013} S. Rohart and A. Thiaville, Phys. Rev. B \textbf{88}, 184422 (2013).

\bibitem{Leonov_EPJ2014} A. O. Leonov, U. K. R{\"o}{\ss}ler, and M. Mostovoy, EPJ Web of Conferences \textbf{75}, 5002, (2014).

\bibitem{Beg_PRB2017} M. Beg, M. Albert, M.-A. Bisotti, D. Cort{\'e}s-Ortu{\~n}o, W. Wang, R. Carey, M. Vousden, O. Hovorka, C. Ciccarelli, C. S. Spencer, C. H. Marrows and H. Fangohr, Phys. Rev. B \textbf{95}, 14433, (2017).

\bibitem{Carey_APL2016} R. Carey, M. Beg, M. Albert, M.-A. Bisotti, D. Cort{\'e}s-Ortu{\~n}o, M. Vousden, W. Wang, O. Hovorka and H. Fangohr, Appl. Phys. Lett. 109, \textbf{122401}, (2016).

\bibitem{Pepper_JAP2018} R. A. Pepper, M. Beg, D. Cort{\'e}s-Ortu{\~n}o, T. Kluyver, M.-A. Bisotti, R. Carey, M. Vousden, M. Albert, W. Wang, O. Hovorka and H. Fangohr, J. Appl. Phys. \textbf{123}, 093903, (2018).

\bibitem{Zang_PRL2011} J. Zang, M. Mostovoy, J.~H. Han and N. Nagaosa, Phys. Rev. Lett. \textbf{107}, 136804 (2011).

\bibitem{Wanjun_NPHYS2017} W. Jiang, X. Zhang, G. Yu, W. Zhang, X. Wang, M. Benjamin~Jungfleisch, J.~E. Pearson, X. Cheng, O. Heinonen, K.~L. Wang, Y. Zhou, A. Hoffmann and S.~G.~E. Velthuiste, Nat. Phys. \textbf{13}, 162 (2017).

\bibitem{Litzius_NPHYS2017} K. Litzius, I. Lemesh, B. Kruger, P. Bassirian, L. Caretta, K. Richter, F. Buttner, K. Sato, O.~A. Tretiakov, J. Forster, R.~M. Reeve, M. Weigand, I. Bykova, H. Stoll, G. Schutz, G.~S.~D. Beach and M. Klaui, Nat. Phys. \textbf{13}, 170 (2017).

\bibitem{Walliser_PRB2000} H. Walliser and G. Holzwarth, Phys. Rev. B \textbf{61}, 2819 (2000).

\bibitem{Kong_PRL2013} L. Kong and J. Zang, Phys. Rev. Lett. \textbf{111}, 067203 (2013).

\bibitem{Lin_PRL2014} S.-Z. Lin, C. D. Batista, C. Reichhardt and A. Saxena, Phys. Rev. Lett. \textbf{112}, 187203 (2014).

\bibitem{Iwasaki_PRB2014} J. Iwasaki, A.~J. Beekman and N. Nagaosa, Phys. Rev. B \textbf{89}, 064412 (2014).

\bibitem{Xichao_NANOTECH2015} X. Zhang, M. Ezawa, D. Xiao, G.~P. Zhao, Y. Liu and Y. Zhou, Nanotechnology \textbf{26}, 225701 (2015).

\bibitem{Xichao_NJP2017} X. Zhang, J. M{\"u}ller, J. Xia, M. Garst, X. Liu and Y. Zhou, New J. Phys. \textbf{19}, 065001 (2017).

\bibitem{OOMMF} M.~J. Donahue and D.~G. Porter, Interagency Report NO. NISTIR 6376, National Institute of Standards and Technology, Gaithersburg, MD (1999) [http://math.nist.gov/oommf/].

\bibitem{Gilbert_1955} T.~L. Gilbert, Phys. Rev. \textbf{100}, 1243 (1955).

\bibitem{Koshibae_NCOMMS2016} W. Koshibae and N. Nagaosa, Nat. Commun. \textbf{7}, 10542 (2016).

\bibitem{Schutte_PRB2014A} C. Sch{\"u}tte and M. Garst, Phys. Rev. B \textbf{90}, 094423 (2014).

\bibitem{Schutte_PRB2014B} C. Sch{\"u}tte, J. Iwasaki, A. Rosch and N. Nagaosa, Phys. Rev. B \textbf{90}, 174434 (2014).

\bibitem{Thiele_PRL1973} A.~A. Thiele, Phys. Rev. Lett. \textbf{30}, 230 (1973).

\end{thebibliography}
\end{document}